\documentclass[twocolumn]{pasj00}
\usepackage[dvips]{color}
\usepackage{multicol}

\begin{document}
\SetRunningHead{N. Narita et al.}{The Rossiter--McLaughlin Effect of TrES-1}
\Received{2007/2/27}
\Accepted{2007/4/9}

\title{Measurement of the Rossiter--McLaughlin Effect\\
in the Transiting Exoplanetary System TrES-1$^{*}$}


\author{
Norio \textsc{Narita},\altaffilmark{1}$^{**}$
Keigo \textsc{Enya},\altaffilmark{2}
Bun'ei \textsc{Sato},\altaffilmark{3}
Yasuhiro \textsc{Ohta},\altaffilmark{1}
Joshua N.\ \textsc{Winn},\altaffilmark{4}
Yasushi \textsc{Suto},\altaffilmark{1,8}
\\
Atsushi \textsc{Taruya},\altaffilmark{1,8}
Edwin L.\ \textsc{Turner},\altaffilmark{5}
Wako \textsc{Aoki},\altaffilmark{6}
Motohide \textsc{Tamura},\altaffilmark{6}
Toru \textsc{Yamada},\altaffilmark{6} and
Yuzuru \textsc{Yoshii}\altaffilmark{7,8}
}

\bigskip

\altaffiltext{1}{
Department of Physics, School of Science, The University of Tokyo,
7-3-1 Hongo, Bunkyo-ku, Tokyo 113--0033
}
\email{narita@utap.phys.s.u-tokyo.ac.jp}

\altaffiltext{2}{
Department of Infrared Astrophysics,
Institute of Space and Astronautical Science,\\
Japan Aerospace Exploration Agency,
3-1-1 Yoshinodai, Sagamihara, Kanagawa 229--8510
}

\altaffiltext{3}{
Okayama Astrophysical Observatory,
National Astronomical Observatory of Japan,\\
3037-5 Honjo, Kamogata, Asakuchi, Okayama 719--0232
}

\altaffiltext{4}{
Department of Physics, and Kavli Institute for Astrophysics
and Space Research,\\
Massachusetts Institute of Technology, Cambridge, MA 02139, USA
}

\altaffiltext{5}{
Princeton University Observatory, Peyton Hall,
Princeton, NJ 08544, USA
}

\altaffiltext{6}{
National Astronomical Observatory of Japan, 
2-21-1 Osawa, Mitaka, Tokyo 181--8588
}

\altaffiltext{7}{
Institute of Astronomy, School of Science,
The University of Tokyo,
2-21-1 Osawa, Mitaka, Tokyo 181--0015
}

\altaffiltext{8}{
Research Center for the Early Universe, The University of Tokyo,
7-3-1 Hongo, Bunkyo-ku, Tokyo 113--0033
}


\maketitle

\begin{abstract}

We report a measurement of the Rossiter--McLaughlin effect in
the transiting extrasolar planetary system TrES-1,
via simultaneous spectroscopic and photometric observations with
the Subaru and MAGNUM telescopes.
By modeling the radial velocity anomaly that was observed during a
transit, we determine the sky-projected angle between the stellar
spin axis and the planetary orbital axis to be
$\lambda = 30 \pm 21$ [deg].
This is the third case for which $\lambda$ has been measured in a transiting
exoplanetary system, and the first demonstration that such measurements
are possible for relatively faint host stars ($V \sim 12$, as compared
to $V \sim 8$ for the other systems).
We also derive a time of mid-transit, constraints on
the eccentricity of the TrES-1b orbit
($e = 0.048 \pm 0.025$),
and upper limits on the mass of the Trojan companions
($\lesssim$14~$M_{\oplus}$) at the 3$\sigma$ level.

\end{abstract}
\footnotetext[*]{Based in part on data collected at Subaru Telescope,
which is operated by the National Astronomical Observatory of Japan.}
\footnotetext[**]{JSPS Fellow.}

\section{Introduction}

The Rossiter--McLaughlin effect (hereafter the RM effect)
is a phenomenon originally reported as a
``rotational effect'' in eclipsing binary systems by
\citet{1924ApJ....60...15R} (for the Beta Lyrae system) and
\citet{1924ApJ....60...22M} (for the Algol system).
In the context of extrasolar planetary science,
the RM effect is seen as a radial velocity anomaly during a planetary
transit caused by the partial occultation of the rotating stellar disk
(see \cite{2005ApJ...622.1118O}, \cite{2006ApJ...650..408G},
or \cite{2007ApJ...655..550G}, for theoretical descriptions).
The radial velocity anomaly depends on the trajectory
of the planet across the disk of the host star, and in particular
on the alignment between that trajectory and the rotation field of the star.
By monitoring this anomaly throughout a transit one can determine whether
or not the planetary orbital axis is well-aligned with the stellar spin axis.
In the Solar System, the orbits of all 8 planets are known to be
well-aligned with the solar equator, but this is not necessarily the case for
exoplanetary systems, or for hot Jupiters in particular.
The key parameter is the sky-projected angle between the stellar
spin axis and the planetary orbital axis, $\lambda$,
and measurements of this ``misalignment angle'' for various
exoplanetary systems will help to place the Solar System in
a broader context.

Specifically, measurements of the RM effect for exoplanetary systems
are important because of the implications for theories of
migration and hot Jupiter formation.
So far, measurements of $\lambda$ for two systems
have been reported; \citet{2000A&A...359L..13Q} and
\citet{2005ApJ...631.1215W} for HD~209458,
\citet{2006ApJ...653L..69W} for HD~189733.
In both of those systems, the host star
is very bright ($V\sim 8$), facilitating the measurement.
The observed values of $\lambda$ for the two systems are small or
consistent with zero,
which would imply that the standard migration mechanism
(planet-disk interaction) does not alter the spin-orbit alignment grossly
during the planetary formation epoch.
However, just these few examples are not enough for statistical constraints
on other hot Jupiter formation theories,
including planet-planet interaction \citep{1996Sci...274..954R,
1996Natur.384..619W}, the ``jumping Jupiter'' model
\citep{2002Icarus...156.570},
or the Kozai mechanism \citep{2003ApJ...589..605W},
which may lead hot Jupiters to have significantly misaligned orbits.
Thus further measurements of the RM effect
for other transiting systems are valuable.
Given that most of ongoing transit surveys target relatively
faint ($V \sim 12$) host stars, it is also important to extend the reach
of this technique to fainter stars.
Further observations for new targets would be useful to constrain
planet formation theories, and more importantly, have a potential to
discover large spin-orbit misalignments, which would be a challenge
to some theoretical models.

In this paper, we report the measurement of the RM effect and
the constraint on $\lambda$ for TrES-1 ($V = 11.8$)
which has a significantly fainter host star than
those in previous studies ($V \sim 8$).
In addition to the fainter host star, this work differs from previous
studies of the RM effect in that we have conducted simultaneous
spectroscopic and photometric observations.
This new strategy offers several potentially important advantages.
First, the simultaneous photometry eliminates any uncertainty
in the results due to the orbital ephemeris and the transit depth.
Although this did not turn out to be crucial for the present work,
it will be useful for newly discovered targets which still have
uncertainty in the times of transits.
Second, the transit depth might be expected to vary due to star spots or
transient events, and indeed evidence for star spots was reported
in HST/ACS photometry for this system \citep{2007prpl.conf..701C}.
Thus simultaneous monitoring is useful to assess anomalies in the
transit depth.
Finally, obtaining all of the data on a single night is useful
to avoid systematic
errors in radial velocities from long-term instrumental instabilities.
Moreover, although it need not be simultaneous, the photometry
also helps to determine the limb-darkening parameter for the visual band,
which can be used in the interpretation of the RM-affected spectra.
In this way, we can determine the limb-darkening parameter of the host
star directly from the data,
instead of assuming a value based on stellar atmosphere models.

We describe our observations in Section 2 and report
the results in Section 3.
Section 4 provides a discussion and summary of this paper.

\section{Observations and Data Reduction}

We observed the planet-hosting K0V star TrES-1 on UT 2006 June 21,
the night of a predicted planetary transit,
using the Subaru 8.2~m Telescope at Mauna Kea and the MAGNUM 2~m Telescope
at Haleakala, both in Hawaii.
The event is predicted as the 238th transit from the first discovery,
namely $E = 238$ ($E$: integer) in the ephemeris by
\citet{2004ApJ...613L.153A};
\begin{eqnarray}
T_c (E) &=& 2453186.8060(\pm 0.0002) \nonumber\\
        &+& E \times 3.030065(\pm 0.000008)
\,\,\, [\textrm{HJD}].
\end{eqnarray}
The transit occurred shortly after midnight. We observed TrES-1 during
5 hours bracketing the predicted transit time, through air masses
ranging from 1.0 to 1.3.

\subsection{MAGNUM Photometry}

The MAGNUM 2~m telescope is located near the Haleakala summit on
the Hawaiian Island of Maui
\citep{1998SPIE.3352..120K,2002RESCEU...235Y,2003AAS...202.3803Y}.
The MAGNUM photometric observation was conducted in parallel
with the Subaru spectroscopic observation.
We employed the Multi-color Imaging Photometer (MIP) using a
$1024 \times 1024$ pixel CCD with a $V$ band filter, covering 4750~\AA\
$< \lambda <$ 6180~\AA, and we set 9 dithering positions (3$\times3$
positions) on the CCD.
The MIP has a $1.5 \times 1.5$ arcmin$^2$ field of view (FOV)
with a pixel scale of 0.277 arcsec/pixel.
We used 2MASS J19041058+3638409 as our
comparison star for differential photometry.
This star is close enough to fit in the MIP field of view,
and is known to be photometrically stable
at a level sufficient for our study (e.g., \cite{2005ApJ...626..523C}).
The exposure time was either 40 or 60 seconds according to observing
conditions so that the photon counts are close to the saturation level
of the CCD, with a readout/setup time of 60 seconds.

We then reduced the images with the standard MIP pipeline
described in \citet{2004ApJ...600L..35M}.
We determined the apparent magnitudes of TrES-1 and the comparison star
using an aperture radius of 20 pixels.
The typical FWHM of each star ranged from 1.4 to 1.9 arcsec
(from 5 to 7 pixels).
We estimated the sky background level with an annulus from 20 to 25
pixels in radius centered on each star, and subtracted the estimated
sky contribution from the aperture flux.
Then we computed the differential magnitude between TrES-1 and the
comparison star.
After these steps, we decorrelated the differential magnitude from
the dithering positions and
eliminated apparent outliers from the light curve,
most of which were obtained at the 9th dithering position.
We do not find any clear correlations with other observing parameters.

For the analysis of transit photometry,
\citet{2006MNRAS.373..231P} studied the time-correlated noise (the
so-called ``red noise'') in detail, and \citet{2006A&A...459..249G}
introduced a simple and useful method to account at least
approximately for the effect of the red noise on parameter
estimation. Based on these studies, we used the following procedure
to determine the appropriate data weights for the MAGNUM photometry
(which are similar to that employed by \cite{2006astro.ph.12224W}).

We first fitted the MAGNUM light curve
with the analytic formula given in \citet{2005ApJ...622.1118O} and
found the residuals between the data and the best-fitting model. Using
only the Poisson noise
as an estimate of the error in each photometric
sample, we found $\chi^2/\nu_{dof} \sim 2.8$ ($\nu_{dof}$: degrees of
freedom), implying that the true errors are significantly in excess of
the Poisson noise.  We also found the residuals in the early part
of the night (before $\sim 2453907.91$ [HJD]) to be significantly
larger than those from later in the night.  This larger scatter could
have been caused by shaking of the telescope by the stronger winds
that occurred during the early part of the night.  We thus estimated
error-bars separately for the early part of the night and the later part
of the night, as described below.

First, we rescaled the error bars to satisfy
$\chi^2/\nu_{dof} = 1.0$ (step 1), namely
0.00259 for the early data and 0.00189 for the late data.
The light curve with these rescaled error bars
is shown in the upper panel of Fig.~\ref{ourdata}.
Next, in order to assess the size of
time-correlated noise for the MAGNUM data,
we solved the following equations,
\begin{eqnarray}
\sigma_1^2 = \sigma_w^2 + \sigma_r^2,\\
\sigma_N^2 = \frac{\sigma_w^2}{N} + \sigma_r^2,
\end{eqnarray}
where $\sigma_1$ is the standard deviation of each residual and
$\sigma_N$ is the standard deviation of the average of the successive
$N$ points.
$\sigma_w$ is called the white noise, which is uncorrelated noise that
averages down as $(1/N)^{1/2}$, while $\sigma_r$ is called the red
noise, which represents correlated noise that remains constant for
specified $N$.
We calculated $\sigma_w$ and $\sigma_r$ for the choice $N=30$
(corresponding to one hour), finding
$\sigma_w = 0.00219$ and $\sigma_r = 0.00139$
for the early data.
On the other hand, we found $\sigma_r^2 < 0$ (suggesting a smaller
level of the red noise) for the late data.
The choice of $N=30$ is fairly arbitrary; other choices of $N$
between 5 and 50 gave similar results.
We then adjusted the error
bars for the early night by multiplying
$[1+N~(\sigma_r/\sigma_w)^2]^{1/2} \sim 3.6$ (step 2).
We did not
change the error bars for the late night data. We adopted these rescaled
uncertainties for subsequent fitting procedures.

\subsection{Subaru HDS Spectroscopy}
\begin{figure}[pt]
 \begin{center}
  \FigureFile(70.5mm,70.5mm){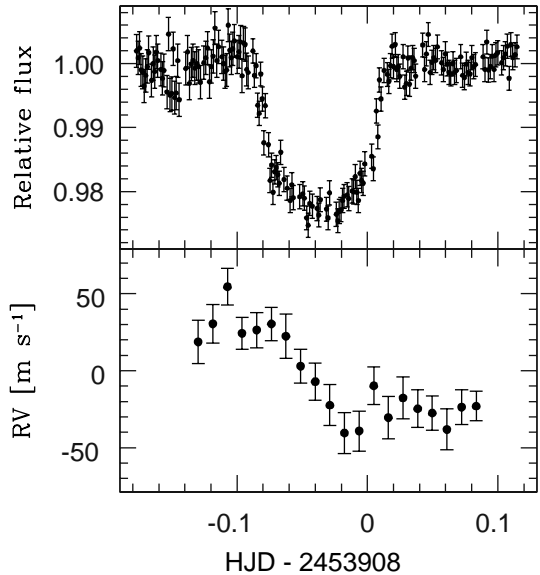}
 \end{center}
  \caption{Top: A photometric
  light curve from the MAGNUM observation (184 samples). The error-bars
  are scaled to satisfy $\chi^2/\nu_{dof} = 1.0$ (see Section 2.1.).
  Bottom: 20 radial velocity samples computed from the Subaru/HDS spectra.
  The values and uncertainties are presented in Table 1.
  \label{ourdata}}
\end{figure}

\begin{table}[pthb]
\caption{Radial velocities obtained with the Subaru/HDS.}
\begin{center}
\begin{tabular}{lcc}
\hline
Time [HJD]  & Value [m~s$^{-1}$] & Error [m~s$^{-1}$]\\
\hline
2453907.87018 &	18.7 &	14.0 \\
2453907.88139 &	30.5 &	12.5 \\
2453907.89262 &	54.6$^a$ &	12.0 \\
2453907.90384 &	24.3 &	10.4 \\
2453907.91506 &	26.4 &	11.4 \\
2453907.92628 &	30.4 &	10.9 \\
2453907.93750 &	22.4 &	14.3 \\
2453907.94873 &	2.9 &	11.0 \\
2453907.95996 &	-7.1 &	12.1 \\
2453907.97119 &	-22.3 &	13.3 \\
2453907.98241 &	-40.5 &	13.3 \\
2453907.99364 &	-39.2 &	13.0 \\
2453908.00488 &	-9.8 &	12.2 \\
2453908.01610 &	-30.5 &	13.8 \\
2453908.02732 &	-17.7 &	13.6 \\
2453908.03854 &	-24.7 &	12.2 \\
2453908.04978 &	-27.5 &	11.1 \\
2453908.06100 &	-38.2 &	13.3 \\
2453908.07223 &	-23.7 &	11.2 \\
2453908.08345 &	-23.0 &	9.6 \\
\hline
\multicolumn{3}{l}{\hbox to 0pt{\parbox{80mm}{\footnotesize
\footnotemark[a]:\,A possible outlier.
}\hss}}
\end{tabular}
\label{rvsummary}
\end{center}
\end{table}

We used the High Dispersion Spectrograph (HDS) on the Subaru Telescope
\citep{2002PASJ...54..855N}.
We employed the standard I2a set-up of the HDS, covering 4940~\AA\
$< \lambda <$ 6180~\AA\, with the Iodine absorption
cell for measuring radial velocities.
The slit width of $0\farcs8$ yielded a spectral resolution of $\sim$ 45000,
and the seeing was between $0\farcs75$ and $1\farcs2$.
The exposure time for TrES-1 was 15 minutes yielding
a typical signal-to-noise ratio (SNR) $\sim$ 60 per pixel.
In order to estimate systematic errors from short term instrumental
variations, we also obtained spectra of the much brighter ($V=4.7$) 
K0V star
HD~185144 before and after the series of TrES-1 exposures.
This star is known to be stable in velocity
\citep{2006ApJ...647..600J}.
We obtained five 30~s exposures of
HD~185144, each having a SNR of approximately 200~pixel$^{-1}$.

We processed the frames with standard IRAF\footnote{The Image
Reduction and Analysis Facility (IRAF) is distributed by the U.S.\
National Optical Astronomy Observatories, which are operated by the
Association of Universities for Research in Astronomy, Inc., under
cooperative agreement with the National Science Foundation.}
procedures and extracted one-dimensional spectra.
Next, we calculated relative radial velocity variations by the algorithm
following \citet{2002PASJ...54..873S}.
We used this algorithm because it properly takes into account the fairly large
changes of the instrumental profile during the observations.
The HDS is known to experience
appreciable instrumental variations even within a single night,
reported in \citet{2004PASJ...56..655W} and \citet{2005PASJ...57..471N}.
We estimated internal errors of the radial velocities from the scatter of
the radial velocity solutions for 2\AA~segments of the spectra.
The typical errors are $10 \sim 15$ [m~s$^{-1}$],
which are reasonable values to be expected from the photon noise limit.
Note that we do not find any evidence of star spots or transient
events during our photometric observation (see Fig.~\ref{ourdata}).
For this reason,
we have not accounted for possible systematic errors in the velocities
due to star spots.
We also reduced the HD~185144 spectra with the same method in order to
check for systematic errors due to short-term instrumental instabilities.
The rms of the radial velocity of HD~185144 is less than 5 [m~s$^{-1}$],
attesting to good instrumental stability.
The resultant radial velocities of TrES-1 and their errors are shown
in Table~\ref{rvsummary} and the lower panel of Fig.~\ref{ourdata}.

\begin{figure*}[phtb]
 \begin{center}
  \FigureFile(120mm,50mm){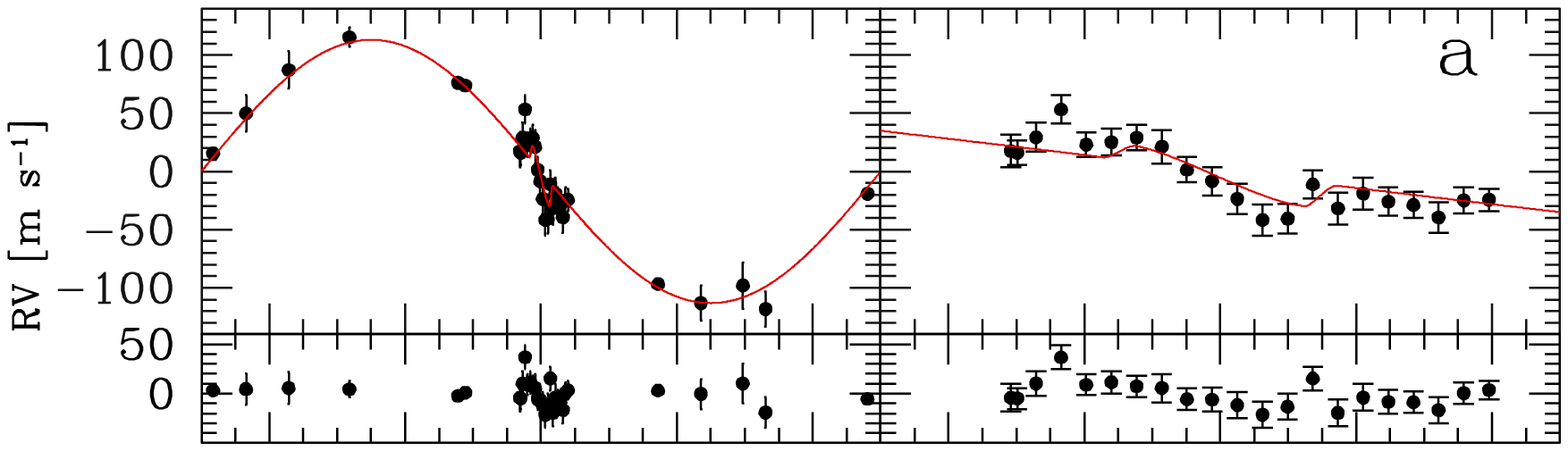}
  \FigureFile(120mm,50mm){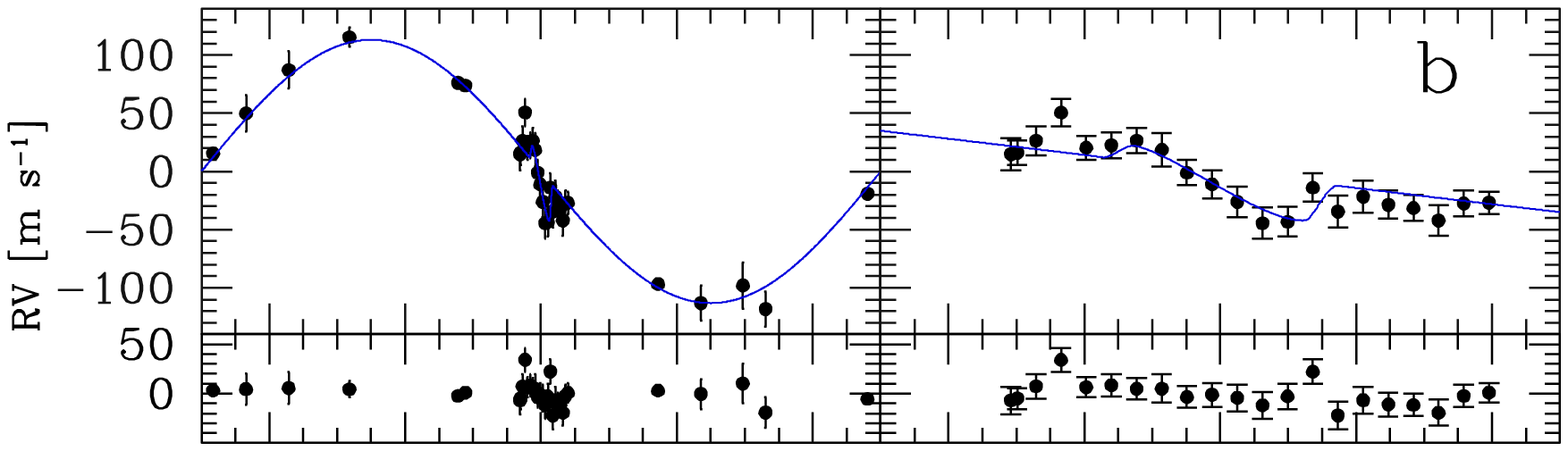}
  \FigureFile(120mm,50mm){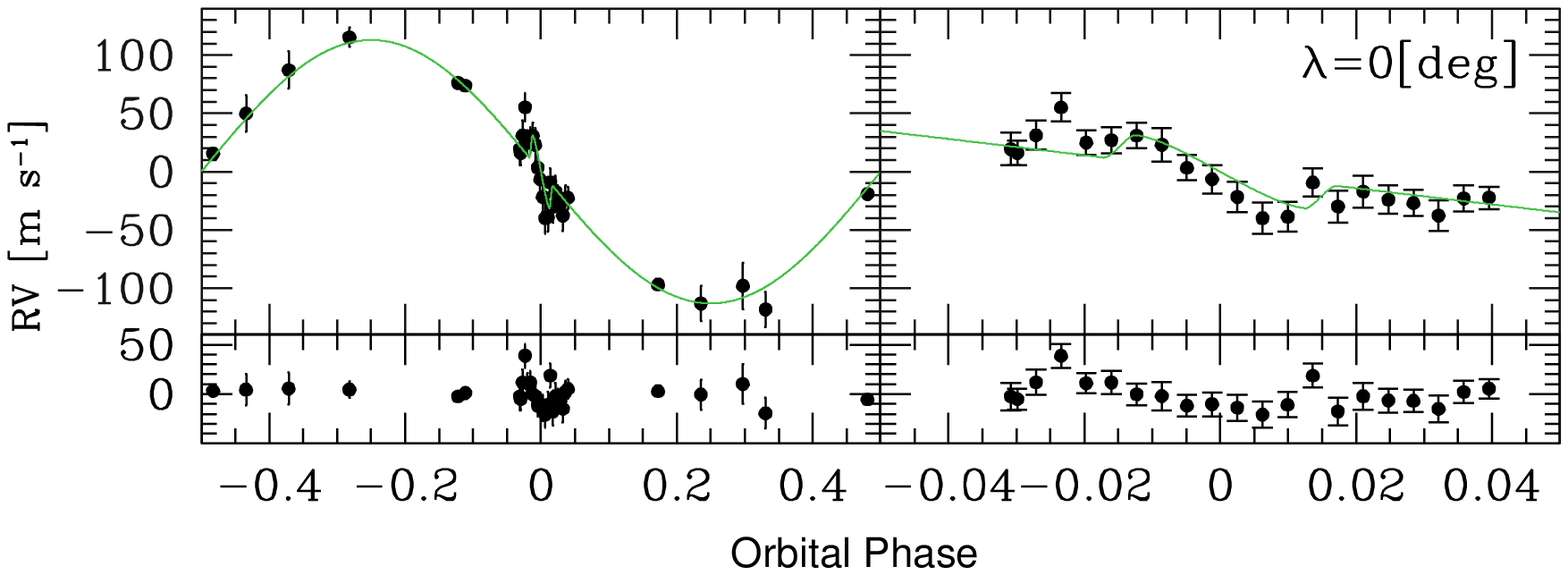}
 \end{center}
  \caption{
  Orbital plots of TrES-1 radial velocities and the best-fitting models,
  phased by $P = 3.0300737$ and $T_c (0) = 2453186.80603$.
  Top (marked with ``a''): With the \textit{a priori} constraint
  on $V \sin I_s$ (see text).
  Middle (marked with ``b''): Without the constraint.
  Bottom (marked with ``$\lambda=0$ [deg]''): Without the constraint and
  assuming that $\lambda=0$ [deg].
  Left panel: A radial velocity plot for the whole orbital phase.
  Right panel: A close-up of the radial velocity plot around
  the transit phase. The waveform around the central
  transit time is caused by the RM effect.
  Bottom panels: Residuals from the best-fit curve.
  \label{rvwithapr}}
 \begin{center}
  \FigureFile(120mm,50mm){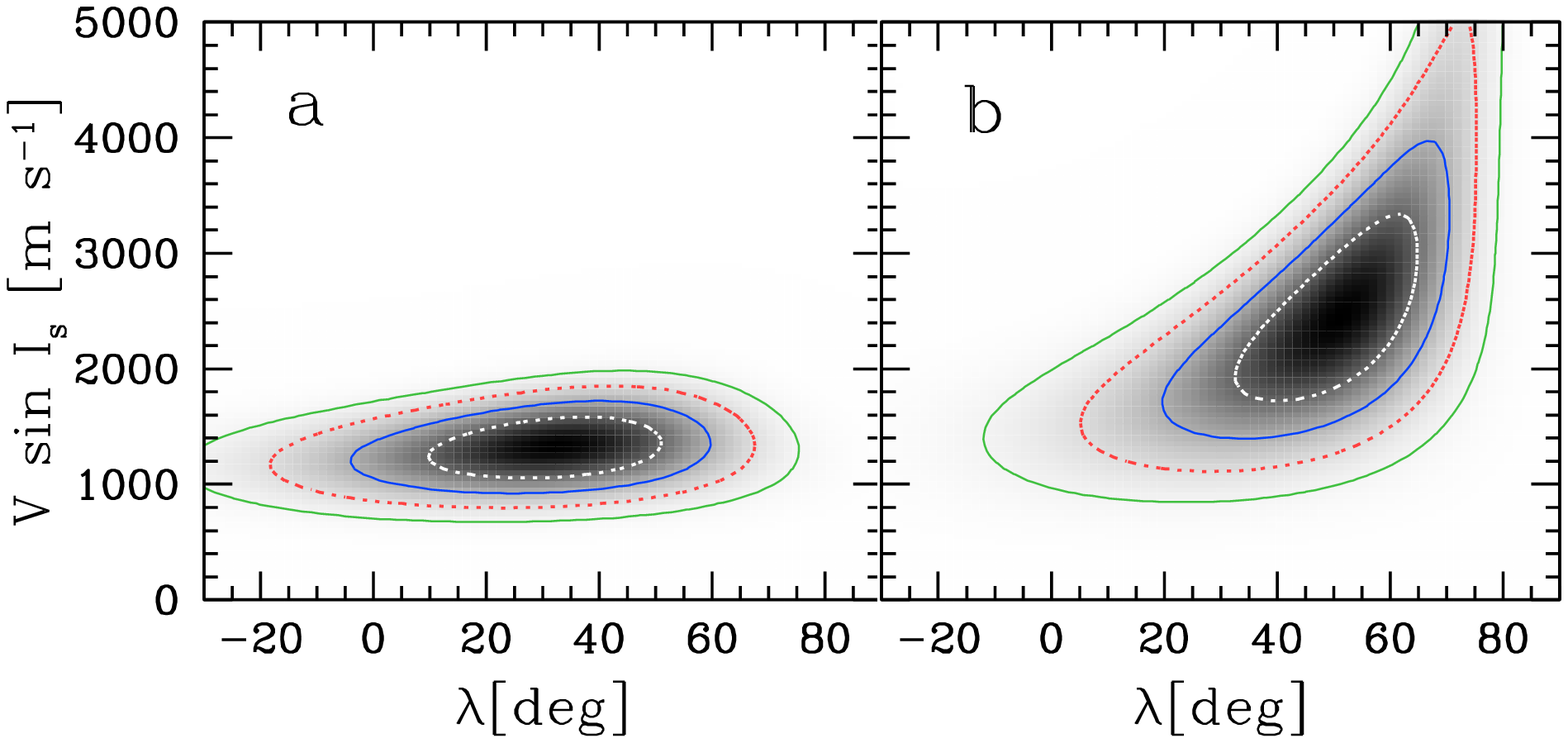} 
 \end{center}
  \caption{Contours of constant $\chi^2$ in
  ($V \sin I_s, \lambda$) space, based on simultaneous fitting
  of 32 radial velocity samples and 1333 photometric samples,
  with (left, marked with a) and without (right, marked with b)
  the \textit{a priori} constraint on $V \sin I_s$.
  The solid line represents $\Delta \chi^2 = 2.30$ (inner)
  and $\Delta \chi^2 = 6.17$ (outer), while
  the dotted line shows $\Delta \chi^2 = 1.00$ (inner)
  and $\Delta \chi^2 = 4.00$ (outer), respectively.}
  \label{vlcontour}
\end{figure*}

\begin{table*}[ptbh]
\caption{Best-fit values and uncertainties$^{\rm{a}}$ of the free parameters.}
\begin{center}
\begin{tabular}{l|cc|cc|cc}
\hline

& \multicolumn{2}{c|}{All data}
& \multicolumn{2}{c|}{All data}
& \multicolumn{2}{c}{Subaru/MAGNUM}\\

& \multicolumn{2}{c|}{With the constraint}
& \multicolumn{2}{c|}{Without the constraint}
& \multicolumn{2}{c}{With the constraint}\\
\hline
Parameter & Value & Uncertainty & Value & Uncertainty & Value & Uncertainty \\
\hline
(All rv samples)
& \multicolumn{2}{c|}{}
& \multicolumn{2}{c|}{}
& \multicolumn{2}{c}{}\\
$K$ [m s$^{-1}$] 
& $113.1$ & $\pm 2.5$
& $113.1$ & $\pm 2.5$
& $115.2^{\rm{b}}$ & fixed \\
$V \sin I_s$ [km s$^{-1}$]
& $1.3$  & $\pm 0.3$ 
& $2.5$  & $\pm 0.8$ 
& $1.3$  & $\pm 0.3$ \\
$\lambda$ [deg] 
& $30$  & $\pm 21$ 
& $48$  & $\pm 17$ 
& $28$  & $\pm 24$ \\
$u_V$ 
& $0.57$ & $\pm 0.05$
& $0.57$ & $\pm 0.05$
& $0.59$ & $\pm 0.05$ \\
$u_z$ 
& $0.37$ & $\pm 0.03$ 
& $0.37$ & $\pm 0.03$
& -- & -- \\
$R_p/R_s$ 
& $0.1382$ & $\pm 0.006$
& $0.1382$ & $\pm 0.006$
& $0.13686^{\rm{c}}$ & fixed \\
$R_s$ [$R_{\odot}$] 
& $0.82$  & $\pm 0.02$
& $0.82$  & $\pm 0.02$
& $0.811^{\rm{c}}$ & fixed \\
$i$ [deg] 
& $88.4$  & $\pm 0.3$
& $88.4$  & $\pm 0.4$
& $88.9^{\rm{c}}$ & fixed \\
$v_1$ [m s$^{-1}$] 
& $1.3$ & $\pm 3.0$
& $4.0$ & $\pm 3.5$
& $0.7$ & $\pm 2.9$ \\
$v_2$ [m s$^{-1}$] 
& $-0.2$ & $\pm 4.9$
& $-0.2$ & $\pm 4.9$
& -- & -- \\
$v_3$ [m s$^{-1}$] 
& $-5.5$ & $\pm 1.6$
& $-5.5$ & $\pm 1.6$
& -- & -- \\
$T_c(234) - 2453000$ [HJD] 
& $895.84298$ & $\pm 0.00015$
& $895.84298$ & $\pm 0.00015$ 
& -- & -- \\
$T_c(235) - 2453000$ [HJD] 
& $898.87342$ & $\pm 0.00014$ 
& $898.87342$ & $\pm 0.00014$
& -- & -- \\
$T_c(236) - 2453000$ [HJD] 
& $901.90371$ & $\pm 0.00016$
& $901.90371$ & $\pm 0.00016$
& -- & -- \\
$T_c(238) - 2453000$ [HJD] 
& $907.96407$ & $\pm 0.00034$
& $907.96407$ & $\pm 0.00034$
& $907.96408$ & $\pm 0.00034$ \\
\hline
(Without the outlier)
& \multicolumn{2}{c|}{}
& \multicolumn{2}{c|}{}
& \multicolumn{2}{c}{}\\
$K$ [m s$^{-1}$]
& $112.8$ & $\pm 2.5$
& $112.8$ & $\pm 2.5$
& $115.2^{\rm{b}}$ & fixed \\
$V \sin I_s$ [km s$^{-1}$]
& $1.3$  & $\pm 0.3$ 
& $2.1$  & $\pm 0.8$ 
& $1.3$  & $\pm 0.3$ \\
$\lambda$ [deg]
& $24$  & $\pm 23$ 
& $39$  & $\pm 21$ 
& $22$  & $\pm 27$ \\
$v_1$ [m s$^{-1}$]
& $1.1$ & $\pm 3.1$
& $0.9$ & $\pm 3.6$
& $-1.5$ & $\pm 3.0$ \\
\hline
\multicolumn{7}{l}{\hbox to 0pt{\parbox{180mm}{\footnotesize
\footnotemark[a]:Computed by $\Delta \chi^2 = 1.00$.
\footnotemark[b]:\citet{2004ApJ...613L.153A}.
\footnotemark[c]:\citet{2006astro.ph.11404W}.
}\hss}}
\end{tabular}
\label{result}
\end{center}
\end{table*}

\section{Results}

As described above, we have obtained 20 radial velocity samples and
184 $V$ band photometric samples taken simultaneously covering the transit.
In addition, in order to search for an optimal solution of orbital
parameters for TrES-1, we incorporate our new data with 12 previously
published radial velocity measurements using the Keck~I telescope
(7 by \cite{2004ApJ...613L.153A} and 5 by \cite{2005ApJ...621.1072L})
and 1149 $z$ band photometric measurements spanning 3 transits
using the FLWO~1.2m telescope \citep{2006astro.ph.11404W}.
The uncertainties of the FLWO data had already been rescaled by the authors
such that $\chi^2/\nu_{dof} = 1.0$ for each transit (namely, the step 1 has
been done). For the step 2, we find $\sigma_r^2 < 0$ for these data, thus
we did not modify these error bars further.
We employ the analytic formulas of radial velocity and photometry
including the RM effect given in \citet{2005ApJ...622.1118O}
and \citet{2006astro.ph.11466O} (hereafter the OTS formulae)
in order to model the observed data.
Note that based on the previous studies by \citet{2005ApJ...631.1215W}
and \citet{2006ApJ...653L..69W}, we have learned that the OTS formulae
systematically underestimate the amplitude of radial velocity
anomaly by approximately 10\%.
This is possibly because the radial velocity
anomaly defined by \citet{2005ApJ...622.1118O}
and that measured by the analysis pipeline are different.
Thus we correct $V \sin I_s$ in the OTS formulas by modifying
$V \sin I_s \textrm{(OTS)} = V \sin I_s \textrm{(real)} * 1.1$.
This correction presumably gives more realistic values for $V \sin I_s$ and
$\lambda$, and has little influence on any of the other parameters.
Here we assume circular orbits of the star and the planet
about the center of mass (namely, $e = 0$). We adopt 
the stellar mass $M_s = 0.87 (\pm 0.03)$ [$M_{\odot}$]
\citep{2005ApJ...621.1072L}
and the orbital period $P = 3.0300737$ [days] and $T_c (0) = 2453186.80603$
[HJD] \citep{2006astro.ph.11404W}.
As a result, our model has 15 free parameters in total.
Eight parameters for the TrES-1 system include the radial velocity
amplitude $K$,
the sky-projected stellar rotational velocity $V \sin I_s$,
the misalignment
angle between the stellar spin and the planetary orbit axes $\lambda$,
the linear limb-darkening parameter for $V$ band $u_V$, the same for
$z$ band $u_z$, the ratio of star-planet radii $R_p/R_s$, the stellar
radius $R_s$, and the orbital inclination $i$.
Here we assume that the limb-darkening parameters for
the spectroscopic and photometric models are the same.\footnote{
This is likely to be a good approximation because the photometric
band is a good match to the region with abundant I$_2$ lines
where the radial velocities are measured.
However, in principle the correspondence is not exact
because the limb-darkening function may not be identical
in the lines as opposed to the continuum, and because
the influence of limb-darkening on the
RV-measuring algorithm has yet to be investigated in detail.
}
We also add three parameters
for velocity offsets to the respective radial velocity dataset $v_{1}$ (for
our template spectrum), $v_{2}$ (for \cite{2004ApJ...613L.153A}) and
$v_{3}$ (for \cite{2005ApJ...621.1072L}),
and four parameters for the times of mid-transit
$T_c (E)\, (E=234, 235, 236, 238)$.

In previous studies of the exoplanetary RM effect,
it was possible and desirable to determine both $V \sin I_s$ and
$\lambda$ from the radial velocity data.  In this case,
there are two reasons to prefer an external determination of $V \sin I_s$.
First, the signal-to-noise ratio of the anomaly is smaller, because
of the faintness of the host star.  This makes it valuable to reduce
the number of degrees of freedom in the model.
Second, the transit geometry is nearly equatorial, which introduces
a very strong degeneracy
between $V \sin I_s$ and $\lambda$, as explained by Gaudi \& Winn (2007).
The alternative we have chosen is to adopt a value for $V \sin I_s$ based
on previous observations,
and use the radial-velocity anomaly to determine $\lambda$.
(We have also investigated our ability to determine both parameters,
as described below.)
\citet{2005ApJ...621.1072L} reported
$V \sin I_s = 1.08 \pm 0.30$
[km~s$^{-1}$] for the TrES-1 host star from
their analysis of the observed spectral line profiles;
this is the most reliable estimate for $V \sin I_s$ to date.
We incorporate this information into our model by adding a term
$\left[ \frac{V \sin I_s - 1.08}{0.30} \right]^2$ to
the $\chi^2$ fitting statistic.  Thus our $\chi^2$ statistic is
\begin{eqnarray}
\chi^2 &=& \sum_{i=1}^{N_{rv}=32} \left[ \frac{v_{i,obs}-v_{i,calc}}
{\sigma_{i}} \right]^2 + \sum_{j=1}^{N_{f}=1333} \left[ \frac{f_{j,obs}
-f_{j,calc}}{\sigma_{j}} \right]^2 \nonumber\\
&+& \left[ \frac{V \sin I_s - 1.08}{0.30} \right]^2,
\end{eqnarray}
where $v_{calc}$ and $f_{calc}$ represent the values calculated by
the OTS formulae with the above parameters.
We find optimal parameters by minimizing the $\chi^2$ statistic of Eq.~(4)
using the AMOEBA algorithm \citep{1992nrca.book.....P},
and estimate confidence levels of the parameters using $\Delta \chi^2$
from the optimal parameter set.
To assess the dependence of our results on the {\it a priori} constraint
on $V \sin I_s$, we also compute and compare the best-fit values and
uncertainties by using another function:
\begin{equation}
\chi^2 = \sum_{i=1}^{N_{rv}=32} \left[ \frac{v_{i,obs}-v_{i,calc}}
{\sigma_{i}} \right]^2 + \sum_{j=1}^{N_{f}=1333} \left[ \frac{f_{j,obs}
-f_{j,calc}}{\sigma_{j}} \right]^2,
\end{equation}
for reference.
In addition, we note that the third radial velocity sample of our data
($t = 2453907.89262$ [HJD]) may appear to be an outlier, but it lies just
about 3$\sigma$ from a theoretical radial velocity curve
(e.g., Fig.~\ref{rvwithapr}).
For clarity, we calculate Eq. (4) and Eq. (5) with and without that sample.

The results for both $\chi^2$ statistics are presented in Table~\ref{result}.
The minimum $\chi^2$ is 1308.57 (1296.39)
for Eq. (4) and 1305.18 (1298.57) for Eq. (5)
with 1350 (1349) degrees of freedom,
where the numbers in parentheses refer to the case without the outlier.
In Fig.~\ref{rvwithapr}, we present the radial velocities and
the best-fit curve with (the top figure, marked with ``a'') and
without (the middle figure, marked with ``b'') the \textit{a priori}
constraint. In addition, we also compute the best-fit curve without
the constraint, but assuming that $\lambda=0$ [deg] (the bottom figure,
marked with ``$\lambda=0$  [deg]'').
Fig.~\ref{vlcontour} plots ($V \sin I_s, \lambda$) contours calculated
with (the left panel, marked with ``a'') and without (the right panel,
marked with ``b'') the \textit{a priori} constraint.
Note that we only show here the results with the possible outlier
in Fig.~\ref{rvwithapr} and Fig.~\ref{vlcontour}, since
the same figures but without the outlier have basically
similar appearance and have less information to show.
As a result, we find $\lambda = 30 \pm 21 \,\,\, (24 \pm 23)$ [deg] and
$V \sin I_s = 1.3 \pm 0.3 \,\,\, (1.3 \pm 0.3)$ [km~s$^{-1}$] for TrES-1
with the \textit{a priori} constraint, and our findings except for $\lambda$
are in good agreement with previous studies
\citep{2004ApJ...613L.153A, 2005ApJ...621.1072L, 2006astro.ph.11404W}.
On the other hand, the result without the \textit{a priori} constraint,
$\lambda = 48 \pm 17 \,\,\, (39 \pm 21)$ [deg] and
$V \sin I_s = 2.5 \pm 0.8 \,\,\, (2.1 \pm 0.8)$ [km~s$^{-1}$], agrees with
the above result within about 1$\sigma$ for $\lambda$ and $V \sin I_s$,
and is also consistent with the previous results for other parameters.
In case we calculate $\chi^2$ without the constraint
but assuming that $\lambda=0$ [deg], we find that the minimum $\chi^2$ is
1309.85 (1298.91) with 1351 (1350) degrees of freedom,
and $V \sin I_s = 1.5 \pm 0.5 \,\,\, (1.5 \pm 0.5)$ [km~s$^{-1}$].
Consequently, our results for $\lambda$ have fairly large uncertainties.
We find at least that the orbital motion of TrES-1b is prograde.
Additional radial velocity measurements during transits, and a
more precise measurement of $V \sin I_s$, would be desirable
to pin down $\lambda$ and help to discriminate between different
modes of migration.

\begin{figure*}[t]
\begin{center}
\begin{minipage}{85mm}
 \begin{center}
  \FigureFile(75mm,75mm){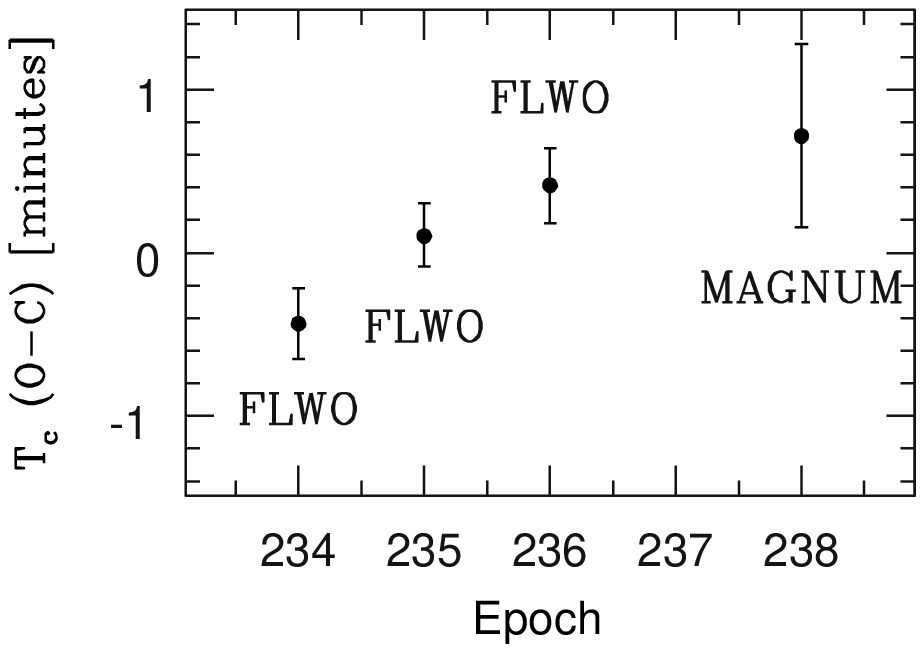}
 \end{center}
  \caption{Timing residuals of the
  time of mid-transit for each epoch ($E$),
  based on the ephemeris of \citet{2006astro.ph.11404W}.
  The results for $E=234, 235, 236$ are determined by the FLWO photometry,
  while $E=238$ is computed by the MAGNUM data.
  \label{tc}}
\end{minipage}
\begin{minipage}{85mm}
 \begin{center}
  \FigureFile(60mm,60mm){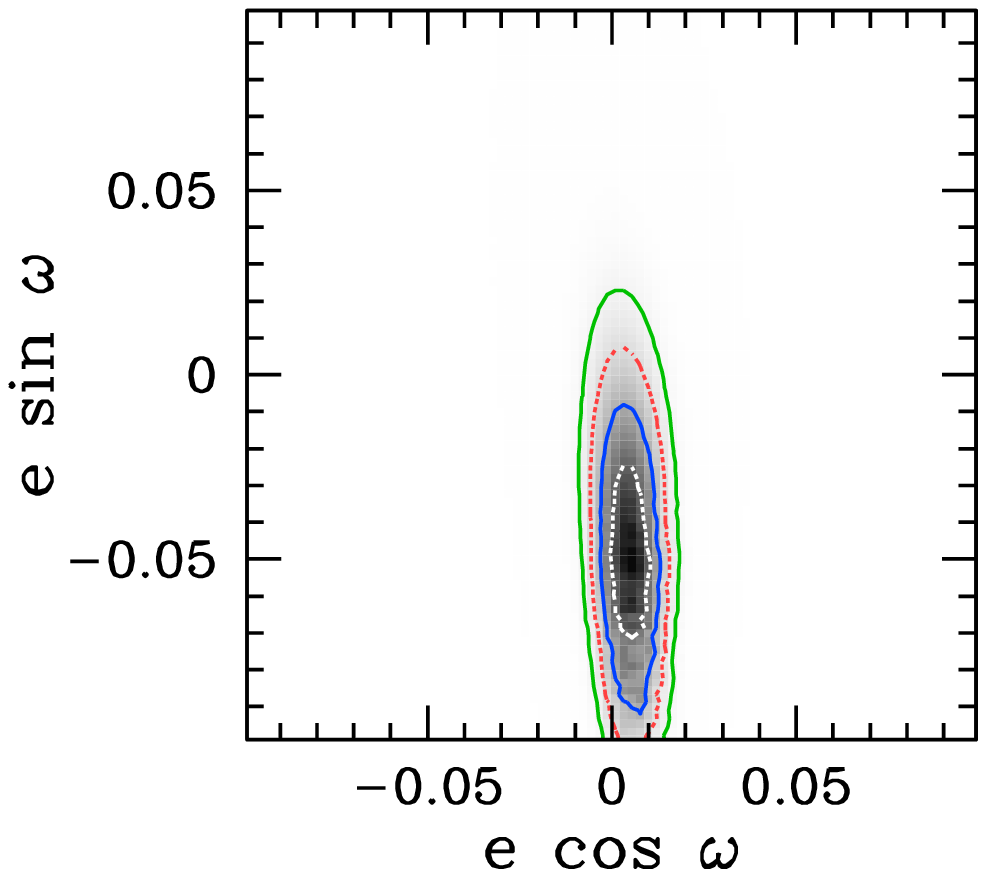}
 \end{center}
  \caption{Contours of constant $\chi^2$ in ($e \cos \omega,
  e \sin \omega$) space, based on simultaneous fitting of 32 radial
  velocity samples and 1333 photometric samples.
  At each grid point, all of the parameters
  except for ($e, \omega$) were optimized.
  Contour line types are the same as those in Fig.~\ref{vlcontour}}
  \label{ewcontour}
\end{minipage}
\end{center}
\end{figure*}

\section{Discussion and Summary}

We have presented simultaneous spectroscopy and photometry of a transit
of TrES-1b, exhibiting a clear detection of the Rossiter-McLaughlin
effect and consequent constraints on the alignment angle between
the stellar spin and the planetary orbital axes.
Our philosophy has been to use
all of the best data available at present.
However, it is also interesting to examine how well we are able to
determine the system parameters using {\it only} the data gathered
on a single night using Subaru and MAGNUM. This is because for future
studies of newly
discovered transiting exoplanetary systems,
higher-precision data from other observatories may not be available.
We repeat the fitting procedure without the Keck and FLWO data but still
assuming the system parameters other than $V \sin I_s$, $\lambda$, $u_V$,
$v_1$, and $T_c (238)$ to be the values presented
in \citet{2004ApJ...613L.153A} and \citet{2006astro.ph.11404W}.
We find almost the same values and uncertainties
for the parameters above (the right side of Table~\ref{result}) as before,
indicating that a single night's data would
have done almost as well as the full data set.

Using transit timing of the TrES-1 system, \citet{2005MNRAS.364L..96S}
reported a constraint on the existence of additional planets in the system.
Subsequently, \citet{2006astro.ph.11404W}
pointed out that $T_c (E)\, (E=234, 235, 236)$ are consistent with
their ephemeris at about the 2$\sigma$ level,
but occurred progressively later than expected.
Here we present $T_c (E)\, (E=238)$ in Fig.~\ref{tc}.
Our result is also consistent with the published ephemeris within
1.5$\sigma$, occurring only slightly later than expected from
\citet{2006astro.ph.11404W}.

In addition to producing transit timing variations, any additional
bodies in the TrES-1 system could excite the orbital eccentricity of
the TrES-1b planet, which could be detectable in the radial velocity
measurements of the host star.  Thus it is worthwhile to put empirical
constraints on the orbital eccentricity.  For this purpose, we compute
$\chi^2$ using Eq.~(4) for fixed values of ($e,\, \omega$) over
numbers of grid points to map out the allowed region in ($e,\,
\omega$) space. In fact the most appropriate parameters are $e \cos
\omega$ and $e \sin \omega$ since their uncertainties are uncorrelated
(see Fig.~\ref{ewcontour}).  The resulting constraints are
$e \cos \omega = 0.005 \pm 0.005$ and $e \sin \omega = -0.048 \pm 0.024$.
Our result for
$e\cos\omega$ is similar to that of \citet{2005ApJ...626..523C}, which
was based on the timing of the secondary eclipse. The value of
$e\cos\omega$ is consistent with zero within 1$\sigma$, and the value
of $e\sin\omega$ is consistent with zero within 2$\sigma$.
Thus we do not find any strong evidence for a nonzero orbital
eccentricity in the TrES-1 system.

There is another interesting application of our data.
Recently, \citet{2006ApJ...652L.137F} studied the detectability of
``hot Trojan'' companions near the L4/L5 points of transiting hot Jupiters,
through any observed difference between the time of vanishing stellar
radial velocity variation ($T_0$)
and the time of the midpoint of the photometric transit ($T_c$).
Our strategy of simultaneous spectroscopic and photometric
transit observations is ideally suited for
searching for the hot Trojan companions.
For the TrES-1 system, \citet{2006ApJ...652L.137F} set an upper limit
on the mass of the Trojan companions
$\simeq 51 M_{\oplus}$ at the 3$\sigma$ level (assuming the circular orbit),
using the radial velocity samples of \citet{2004ApJ...613L.153A}.
Here we compute $\Delta t = T_0 - T_c (238)$ using all available
out-of-transit radial velocity samples with (without) the possible outlier,
and both for the circular and the eccentric orbit.
We find
$\Delta t =  3.2 \pm 11.2 \,\, (3.2 \pm 11.8)$ [min] (circular) and
$\Delta t =  33.0 \pm 52.6 \,\, (33.7 \pm 44.8)$ [min] (eccentric).
Accordingly, we set constraints on the mass of the Trojan companions,
$M_T$, which is defined as the difference in the mass
at L4 ($M_{T,L4}$) and the mass at L5 ($M_{T,L5}$)
(namely, $M_T \equiv M_{T,L4} - M_{T,L5}$), through the relation;
\begin{eqnarray}
M_T &\simeq& \frac{4\pi}{\sqrt{3}} M_p \frac{\Delta t}{P},\\
\sigma_{M_T} &\simeq& \frac{4\pi}{\sqrt{3}} M_p \frac{\sigma_{\Delta t}}{P},
\end{eqnarray}
where $\sigma_{M_T}$ and $\sigma_{\Delta t}$ indicate the uncertainties of
$M_T$ and $\Delta t$, respectively (Eq.~(1) and Eq.~(2) in
\cite{2006ApJ...652L.137F}).
Note that we adopt $M_p = 0.73 \pm 0.03$ [$M_{\mathrm{Jup}}$] which is
determined by this work.
We find
$M_T =  1.2 \pm 4.3 \,\,\, (1.2 \pm 4.6)$ [$M_{\oplus}$] (circular) and
$M_T =  13 \pm 20 \,\,\, (13 \pm 17)$ [$M_{\oplus}$] (eccentric).
As a result, we exclude the Trojan companions near the L4 point more
massive than
$\simeq 14 \,\,\, (15)$ [$M_{\oplus}$] if the orbit is circular, and
$\simeq 74 \,\,\, (65)$ [$M_{\oplus}$] if the orbit is allowed to be
eccentric, both at the 3$\sigma$ level.
Our constraint under the reasonable assumption of a circular orbit is
more stringent than that of \citet{2006ApJ...652L.137F} by a factor of 4,
because we have increased the number of radial velocity samples
and because our data cover the critical phase to determine $T_0$.
Consequently, we conclude that we do not find any sign
of the existence of additional bodies in the TrES-1 system at present.

In this paper, we have placed a constraint on the sky-projected angle
between the stellar spin axis and the planetary orbital axis for the
TrES-1 system, namely $\lambda = 30 \pm 21$ [deg] using all available data
and information from previous studies.
Although we can not discriminate whether the spin-orbit angle in this
system is well-aligned or not at this point, 
our constraint on $\lambda$ clearly indicates the prograde orbital motion
of TrES-1b.
The uncertainty is larger
than in previous studies ($\sim 1$ [deg] for HD~209458 and HD~189733)
because the host star is significantly fainter in this case.  Although
further radial velocity measurements during transit would be necessary
to pin down $\lambda$ more stringently, we have demonstrated for the
first time that such measurements are possible for such a faint
target.  This is important because most of the newly discovered
transiting planets from ongoing transit surveys will have relatively
faint host stars.
For example, the new targets that were discovered in 2006,
namely XO-1 \citep{2006ApJ...648.1228M}, TrES-2
\citep{2006ApJ...651L..61O}, HAT-P-1 \citep{2006astro.ph..9369B},
WASP-1 and WASP-2 \citep{2007MNRAS.tmp.1491C}, are all in this category.
Combining future measurements of $\lambda$ in other
transiting systems, we would be able to determine the distribution of
$\lambda$ for exoplanetary systems with useful statistical accuracy.

\bigskip

We acknowledge close support of our
observations by Akito Tajitsu, who is a Support Scientist for Subaru
HDS.
We appreciate the detailed and helpful critique of the manuscript
by the referee, Scott Gaudi.
We also thank Kazuhiro Yahata, Shunsaku Horiuchi, Takahiro Nishimichi,
and Hiroshi Ohmuro for useful discussions.
N.N. is supported by a Japan Society for Promotion of Science (JSPS)
Fellowships for Research.
This work is supported in part by a Grant-in-Aid for Scientific Research
from the JSPS (No.14102004, 16340053, 17740106),
and MEXT Japan, Grant-in-Aid for Scientific Research on Priority Areas,
"Development of Extra-solar Planetary Science,"
and NASA grant NAG5-13148.
We wish to recognize and acknowledge the very
significant cultural role and reverence that the summit of Mauna Kea
has always had within the indigenous Hawaiian community.


\bigskip


\begin{thebibliography}{}
\expandafter\ifx\csname natexlab\endcsname\relax\def\natexlab#1{#1}\fi

\bibitem[{{Alonso} {et~al.}(2004){Alonso}, {Brown}, {Torres}, {Latham},
  {Sozzetti}, {Mandushev}, {Belmonte}, {Charbonneau}, {Deeg}, {Dunham},
  {O'Donovan}, \& {Stefanik}}]{2004ApJ...613L.153A}
{Alonso}, R., \etal\ 2004, \apjl, 613, L153

\bibitem[{{Bakos} {et~al.}(2006){Bakos}, {Noyes}, {Kovacs}, {Latham},
  {Sasselov}, {Torres}, {Fischer}, {Stefanik}, {Sato}, {Johnson}, {Pal},
  {Marcy}, {Butler}, {Esquerdo}, {Stanek}, {Lazar}, {Papp}, {Sari}, \&
  {Sipocz}}]{2006astro.ph..9369B}
{Bakos}, G.~A., \etal\ 2006, astro-ph/0609369

\bibitem[{{Cameron} {et~al.}(2007){Cameron}, {Bouchy}, {H{\'e}brard}, {Maxted},
  {Pollacco}, {Pont}, {Skillen}, {Smalley}, {Street}, {West}, {Wilson},
  {Aigrain}, {Christian}, {Clarkson}, {Enoch}, {Evans}, {Fitzsimmons},
  {Fleenor}, {Gillon}, {Haswell}, {Hebb}, {Hellier}, {Hodgkin}, {Horne},
  {Irwin}, {Kane}, {Keenan}, {Loeillet}, {Lister}, {Mayor}, {Moutou}, {Norton},
  {Osborne}, {Parley}, {Queloz}, {Ryans}, {Triaud}, {Udry}, \&
  {Wheatley}}]{2007MNRAS.tmp.1491C}
{Cameron}, A.~C., \etal\ 2007, \mnras, 1491

\bibitem[{{Charbonneau} {et~al.}(2005){Charbonneau}, {Allen}, {Megeath},
  {Torres}, {Alonso}, {Brown}, {Gilliland}, {Latham}, {Mandushev}, {O'Donovan},
  \& {Sozzetti}}]{2005ApJ...626..523C}
{Charbonneau}, D., \etal\ 2005, \apj, 626, 523

\bibitem[{{Charbonneau} {et~al.}(2007){Charbonneau}, {Brown}, {Burrows}, \&
  {Laughlin}}]{2007prpl.conf..701C}
{Charbonneau}, D., {Brown}, T.~M., {Burrows}, A., \& {Laughlin}, G. 2007, in
  Protostars and Planets, ed. V, B. Reipurth, D. Jewitt, and K. Keil
  (Tucson: University of Arizona Press), 701

\bibitem[{{Gaudi} \& {Winn}(2007)}]{2007ApJ...655..550G}
{Gaudi}, B.~S. \& {Winn}, J.~N. 2007, \apj, 655, 550

\bibitem[{{Gillon} {et~al.}(2006){Gillon}, {Pont}, {Moutou}, {Bouchy},
  {Courbin}, {Sohy}, \& {Magain}}]{2006A&A...459..249G}
{Gillon}, M., {Pont}, F., {Moutou}, C., {Bouchy}, F., {Courbin}, F., {Sohy},
  S., \& {Magain}, P. 2006, \aap, 459, 249

\bibitem[{{Gim{\'e}nez}(2006)}]{2006ApJ...650..408G}
{Gim{\'e}nez}, A. 2006, \apj, 650, 408

\bibitem[{{Ford} \& {Gaudi}(2006)}]{2006ApJ...652L.137F}
{Ford}, E.~B. \& {Gaudi}, B.~S. 2006, \apjl, 652, L137

\bibitem[{{Johnson} {et~al.}(2006){Johnson}, {Marcy}, {Fischer}, {Laughlin},
  {Butler}, {Henry}, {Valenti}, {Ford}, {Vogt}, \&
  {Wright}}]{2006ApJ...647..600J}
{Johnson}, J.~A., \etal\ 2006, \apj, 647, 600

\bibitem[{{Kobayashi} {et~al.}(1998){Kobayashi}, {Yoshii}, {Peterson},
  {Miyazaki}, {Aoki}, {Minezaki}, {Kawara}, {Enya}, {Okada}, {Suganuma},
  {Greene}, {O'Brien}, \& {Randall}}]{1998SPIE.3352..120K}
{Kobayashi}, Y., \etal\ 1998, in Proc. SPIE Vol.
  3352, Advanced Technology Optical/IR Telescopes VI,
  ed. L.~M. {Stepp}, 120

\bibitem[{{Laughlin} {et~al.}(2005){Laughlin}, {Wolf},
  {Vanmunster}, {Bodenheimer}, {Fischer}, {Marcy}, {Butler}, \&
  {Vogt}}]{2005ApJ...621.1072L}
{Laughlin}, G., {Wolf}, A., {Vanmunster}, T., {Bodenheimer}, P., {Fischer}, D.,
  {Marcy}, G., {Butler}, P., \& {Vogt}, S. 2005, \apj, 621, 1072

\bibitem[{{Marzari} \& {Weidenschilling}(2002)}]{2002Icarus...156.570}
  {Marzari}, F., \& {Weidenschilling}, S.~J.\ 2002, Icarus, 156, 570

\bibitem[{{McCullough} {et~al.}(2006){McCullough}, {Stys}, {Valenti},
  {Johns-Krull}, {Janes}, {Heasley}, {Bye}, {Dodd}, {Fleming}, {Pinnick},
  {Bissinger}, {Gary}, {Howell}, \& {Vanmunster}}]{2006ApJ...648.1228M}
{McCullough}, P.~R., \etal\ 2006, \apj, 648, 1228

\bibitem[{{McLaughlin}(1924)}]{1924ApJ....60...22M}
{McLaughlin}, D.~B. 1924, \apj, 60, 22

\bibitem[{{Minezaki} {et~al.}(2004){Minezaki}, {Yoshii}, {Kobayashi}, {Enya},
  {Suganuma}, {Tomita}, {Aoki}, \& {Peterson}}]{2004ApJ...600L..35M}
{Minezaki}, T., {Yoshii}, Y., {Kobayashi}, Y., {Enya}, K., {Suganuma}, M.,
  {Tomita}, H., {Aoki}, T., \& {Peterson}, B.~A. 2004, \apjl, 600, L35

\bibitem[{{Narita} {et~al.}(2005){Narita}, {Suto}, {Winn}, {Turner}, {Aoki},
  {Leigh}, {Sato}, {Tamura}, \& {Yamada}}]{2005PASJ...57..471N}
{Narita}, N., \etal\ 2005, \pasj, 57, 471

\bibitem[{{Noguchi} {et~al.}(2002){Noguchi}, {Aoki}, {Kawanomoto}, {Ando},
  {Honda}, {Izumiura}, {Kambe}, {Okita}, {Sadakane}, {Sato}, {Tajitsu},
  {Takada-Hidai}, {Tanaka}, {Watanabe}, \& {Yoshida}}]{2002PASJ...54..855N}
{Noguchi}, K., \etal\ 2002, \pasj, 54, 855

\bibitem[{{O'Donovan} {et~al.}(2006){O'Donovan}, {Charbonneau}, {Mandushev},
  {Dunham}, {Latham}, {Torres}, {Sozzetti}, {Brown}, {Trauger}, {Belmonte},
  {Rabus}, {Almenara}, {Alonso}, {Deeg}, {Esquerdo}, {Falco}, {Hillenbrand},
  {Roussanova}, {Stefanik}, \& {Winn}}]{2006ApJ...651L..61O}
{O'Donovan}, F.~T., \etal\ 2006, \apjl, 651, L61

\bibitem[{{Ohta} {et~al.}(2005){Ohta}, {Taruya}, \&
  {Suto}}]{2005ApJ...622.1118O}
{Ohta}, Y., {Taruya}, A., \& {Suto}, Y. 2005, \apj, 622, 1118

\bibitem[{{Ohta} {et~al.}(2006){Ohta}, {Taruya}, \&
  {Suto}}]{2006astro.ph.11466O}
{Ohta}, Y., {Taruya}, A., \& {Suto}, Y. 2006, astro-ph/0611466

\bibitem[{{Pont} {et~al.}(2006){Pont}, {Zucker}, \&
  {Queloz}}]{2006MNRAS.373..231P}
{Pont}, F., {Zucker}, S., \& {Queloz}, D. 2006, \mnras, 373, 231

\bibitem[{{Press} {et~al.}(1992){Press}, {Teukolsky}, {Vetterling}, \&
  {Flannery}}]{1992nrca.book.....P}
{Press}, W.~H., {Teukolsky}, S.~A., {Vetterling}, W.~T., \& {Flannery}, B.~P.
  1992, {Numerical recipes in C. The art of scientific computing} (Cambridge:
  University Press, |c1992, 2nd ed.)

\bibitem[{{Queloz} {et~al.}(2000){Queloz}, {Eggenberger}, {Mayor}, {Perrier},
  {Beuzit}, {Naef}, {Sivan}, \& {Udry}}]{2000A&A...359L..13Q}
{Queloz}, D., {Eggenberger}, A., {Mayor}, M., {Perrier}, C., {Beuzit}, J.~L.,
  {Naef}, D., {Sivan}, J.~P., \& {Udry}, S. 2000, \aap, 359, L13

\bibitem[{{Rasio} \& {Ford}(1996)}]{1996Sci...274..954R}
{Rasio}, F.~A. \& {Ford}, E.~B. 1996, Science, 274, 954

\bibitem[{{Rossiter}(1924)}]{1924ApJ....60...15R}
{Rossiter}, R.~A. 1924, \apj, 60, 15

\bibitem[{{Sato} {et~al.}(2002){Sato}, {Kambe}, {Takeda}, {Izumiura}, \&
  {Ando}}]{2002PASJ...54..873S}
{Sato}, B., {Kambe}, E., {Takeda}, Y., {Izumiura}, H., \& {Ando}, H. 2002,
  \pasj, 54, 873

\bibitem[{{Steffen} \& {Agol}(2005)}]{2005MNRAS.364L..96S}
{Steffen}, J.~H. \& {Agol}, E. 2005, \mnras, 364, L96

\bibitem[{{Weidenschilling} \& {Marzari}(1996)}]{1996Natur.384..619W}
{Weidenschilling}, S.~J. \& {Marzari}, F. 1996, \nat, 384, 619

\bibitem[{{Winn} {et~al.}(2006{\natexlab{a}}){Winn}, {Holman}, {Henry},
  {Roussanova}, {Enya}, {Yoshii}, {Shporer}, {Mazeh}, {Johnson}, {Narita}, \&
  {Suto}}]{2006astro.ph.12224W}
{Winn}, J.~N., \etal\ 2006{\natexlab{a}}, astro-ph/0612224

\bibitem[{{Winn} {et~al.}(2006{\natexlab{b}}){Winn}, {Holman}, \&
  {Roussanova}}]{2006astro.ph.11404W}
{Winn}, J.~N., {Holman}, M.~J., \& {Roussanova}, A. 2006{\natexlab{b}},
  astro-ph/0611404

\bibitem[{{Winn} {et~al.}(2006{\natexlab{c}}){Winn}, {Johnson}, {Marcy},
  {Butler}, {Vogt}, {Henry}, {Roussanova}, {Holman}, {Enya}, {Narita}, {Suto},
  \& {Turner}}]{2006ApJ...653L..69W}
{Winn}, J.~N., \etal\ 2006{\natexlab{c}}, \apjl, 653,
  L69

\bibitem[{{Winn} {et~al.}(2005){Winn}, {Noyes}, {Holman}, {Charbonneau},
  {Ohta}, {Taruya}, {Suto}, {Narita}, {Turner}, {Johnson}, {Marcy}, {Butler},
  \& {Vogt}}]{2005ApJ...631.1215W}
{Winn}, J.~N., \etal\ 2005, \apj, 631, 1215

\bibitem[{{Winn} {et~al.}(2004){Winn}, {Suto}, {Turner}, {Narita}, {Frye},
  {Aoki}, {Sato}, \& {Yamada}}]{2004PASJ...56..655W}
{Winn}, J.~N., {Suto}, Y., {Turner}, E.~L., {Narita}, N., {Frye}, B.~L.,
  {Aoki}, W., {Sato}, B., \& {Yamada}, T. 2004, \pasj, 56, 655

\bibitem[{{Wu} \& {Murray}(2003)}]{2003ApJ...589..605W}
{Wu}, Y. \& {Murray}, N. 2003, \apj, 589, 605

\bibitem[{{Yoshii}(2002)}]{2002RESCEU...235Y}
{Yoshii}, Y. 2002, in New Trends in Theoretical and Observational Cosmology,
  ed. K. Sato \& T. Shiromizu (Tokyo: Universal Academy Press), 235

\bibitem[{{Yoshii} {et~al.}(2003){Yoshii}, {Kobayashi}, \&
  {Minezaki}}]{2003AAS...202.3803Y}
{Yoshii}, Y., {Kobayashi}, Y., \& {Minezaki}, T. 2003, in Bulletin of the
  American Astronomical Society, 752

\end{thebibliography}
\end{document}